\documentclass[11pt,a4paper]{JHEP3}
\usepackage{amssymb} 
\usepackage{euscript}

\def\a{{\alpha}}
\def\b{{\beta}}

\def\0{\nonumber}
\def\one{{\bf 1}}
\def\e{{\bf e}}
\def\det{{\rm Det}}
\def\Y{{Y_\varepsilon}}
\newcommand\ES{\EuScript{S}}
\newcommand\Z{\EuScript{Z}}
\newcommand\T{\EuScript{T}}
\newcommand\A{\EuScript{A}}
\newcommand\B{\EuScript{B}}
\newcommand\X{\EuScript{X}}
\newcommand\V{\EuScript{V}}
\newcommand\EY{\EuScript{Y}}
\newcommand\R{\EuScript{R}}
\newcommand\N{\EuScript{N}}
\newcommand\EP{\EuScript{P}}
\newcommand\W{\EuScript{W}}
\newcommand\I{\mathbb{I}}
\newcommand\ee{\end{eqnarray}}	 	
\newcommand\be{\begin{eqnarray}}
\newcommand\ba{\begin{array}}			
\newcommand\ea{\end{array}}
\newcommand\eeq{\end{equation}}	 	
\newcommand\beq{\begin{equation}}

\preprint{SISSA/21/02/EP\\\tt hep-th/0203188}

\title{Vacuum String Field Theory with B field}

\author{ L.Bonora, D.Mamone, M.Salizzoni\\
International School for Advanced Studies (SISSA/ISAS)\\
Via Beirut 2--4, 34014 Trieste, Italy, and INFN, Sezione di 
Trieste\\
E-mail:   \email{bonora@sissa.it}, \email{mamone@sissa.it}, 
\email{sali@sissa.it} }

\abstract{We continue the analysis of Vacuum String Field Theory 
in the presence of a constant $B$ field. In particular we give a
proof of the ratio of brane tensions is the expected one. 
On the wake of the 
recent literature we introduce wedge--like states and orthogonal 
projections. Finally we show a few examples of the 
smoothing out effects of the $B$ field on some of the singularities
that appear in VSFT.}


\begin{document}
\section{Introduction}

Witten's String Field Theory (SFT), \cite{W1} in the presence 
of a constant background 
$B$ field has been studied in ref. \cite{W2} and \cite{Sch} in the 
field theory limit (see also the final considerations of \cite{MT}).
In this limit the SFT $*$ product factorizes into the ordinary 
Witten $*$ product and the Moyal product.  More generally, 
it was proven by \cite{sugino,KT} that, when a $B$ 
field is switched on, the kinetic term of the SFT action remains 
unchanged while the three string vertex changes, being multiplied by 
a (cyclically invariant) noncommutative phase factor 
(see \cite{sugino,KT}). Starting from this result, in \cite{BMS} 
we began the exploration of the effects of a $B$ field in a 
nonperturbative regime of Vacuum String Field Theory (VSFT). 
In particular we were able to show that exact solutions can be 
written down for tachyonic lumps much in the same way as one 
finds analogous solutions without $B$ field.

In this paper we continue the analysis started in \cite{BMS} and show
that many results previously obtained in the matter sector of VSFT 
without $B$ field (see ref.[8-27]) can be generalized to the VSFT 
with $B$ field. The message we would like to convey in this paper
is that the introduction of a constant background $B$ field is not
a terrible embarrassment in developing SFT. On the other hand, 
it may turn out to be a useful device. In VSFT there arise several
singularities. We have good reasons to believe that the well--known 
smoothing out effects of a $B$ field may help in taming some of them.
At the end of this paper we present 
an example of such beneficial effects due to the $B$ field. 

The paper is organized as follows. In the next section we collect
a series of previously obtained results, which we need in the
sequel of this paper. In section 3 we present a proof that  
$\R=1$, where $\R$ is the ratio introduced in \cite{BMS}. This implies
that the ratio of tensions between the D25--brane and the 
D23--dimensional tachyonic lump is as conjectured in \cite{BMS}. 
In section 4 we show that we can extend without much pain to the case
of nonvanishing $B$ field a series of results which were previously 
derived in the case $B=0$.
In section 5 we analyze the effect of the $B$ field on some of the 
VSFT singularities. We prove that the presence of a $B$
field removes the singularity of the tachyonic lumps. For instance
we find the GMS solitons, \cite{GMS}, and we show that
the singular geometry of the sliver, found in \cite{MT}, is smoothed
out by the $B$ field in the lump solutions. In particular we show 
that the string midpoint is not confined anymore to the hyperplane 
of vanishing transverse dimensions.

\section{Summary of previous results}

Vacuum String Field Theory (VSFT) was defined in \cite{RSZ1}. It
is conjectured to represent SFT on the stable vacuum of Witten's
SFT. The VSFT action has the same form as Witten's SFT action with 
the BRST operator $Q$ replaced by a new one, usually denoted 
${\cal Q}$, which has the characteristics of being universal. 
As a matter of fact 
in \cite{RSZ3}, see also \cite{HK,Oku1,Oku2,RSZ4,RSZ5,Kishi} and
\cite{RSZ3,GT,KO,Moeller},
an explicit representation of ${\cal Q}$ has 
been proposed, purely 
in terms of ghost fields. Now, the equation of motion of VSFT is
\be
{\cal Q} \Psi = - \Psi * \Psi\label{EOM}
\ee
One looks for nonperturbative solutions in the form
\be
\Psi= \Psi_m \otimes \Psi_g\label{ans}
\ee
where $\Psi_g$ and $\Psi_m$ depend purely on ghost and matter 
degrees of freedom, respectively. Then eq.(\ref{EOM}) splits into
\be
 {\cal Q} \Psi_g & = & - \Psi_g * \Psi_g\label{EOMg}\\
\Psi_m & = & \Psi_m * \Psi_m\label{EOMm}
\ee
Eq.(\ref{EOMg}) will not be involved in our analysis since ghosts 
are unaffected by the presence of a $B$ field. Therefore we will 
concentrate on solutions of (\ref{EOMm}). 

The value of the action for such solutions is given by 
\be
{\cal S}(\Psi) = {\EuScript K} \langle\Psi_m|\Psi_m\rangle\label{action2}
\ee
where ${\EuScript K}$ contains the ghost contribution. As shown 
in \cite{RSZ3}, ${\EuScript K}$  is infinite unless it is suitably 
regularized. Nevertheless, as argued in \cite{RSZ3} 
a coupled solution of (\ref{EOMg}) and (\ref{EOMm}), even if it 
is naively singular, is nevertheless a 
representative of the corresponding class of smooth solutions.

In \cite{BMS} we found solutions of eq.(\ref{EOMm})
when a constant $B$ field is turned on along some space 
directions. We consider here only the simplest $B$ field 
configuration, i.e. when $B$ is nonvanishing in the two space 
directions, say the $24$--th and $25$--th ones (see \cite{BMS}
for generalizations). Let us denote these directions with the 
Lorentz indices $\alpha$ and $\beta$.
Then, as is well--known \cite{SW}, in these two direction 
we have a new effective metric $G_{\alpha\beta}$, 
the open string metric, as well as an effective antisymmetric 
parameter $\theta_{\alpha\beta}$. If we set
\be
B_{\a\b}= \left(\matrix {0&B\cr -B&0\cr}\right)\label{B}
\ee
they take the explicit form
\be
G_{\alpha\beta} = \sqrt{{\rm Det G}} \,\delta_{\a\b},\quad\quad
\theta^{\alpha\beta} =  -(2\pi \a')^2  B\epsilon^{\a\b},\quad\quad 
{\rm Det G} = \left( 1+ (2 \pi B)^2\right)^2,  \label{Gtheta}
\ee
where $\epsilon^{\a\b}$ is the $2\times 2$ antisymmetric symbol
with $\epsilon^1{}_2=1$ (notice the slight change of conventions
with respect to \cite{BMS}).  

The presence of the $B$ field modifies the three--string vertex 
only in the 24-th and 
25-th direction, which, in view of the D--brane 
interpretation, we call the transverse ones.
After turning on the $B$--field the three--string vertex becomes 
\be
|V_3 \rangle' = 
|V_{3,\perp}\rangle ' \,\otimes\,|V_{3,\|}\rangle \label{split'}
\ee
$|V_{3,\|}\rangle$ is the same as in the ordinary case 
(without $B$ field), while
\be
|V_{3,\perp}\rangle'= K_2\, e^{-E'}|\tilde 0\rangle_{123}\label{V3'} 
\ee
with
\be
&&K_2= \frac {\sqrt{2\pi b^3}}{A^2 (4a^2+3)}({\rm Det} G)^{1/4},\label{K2}\\
&&E'= \frac 12 \sum_{r,s=1}^3 \sum_{M,N\geq 0} a_M^{(r)\a\dagger}
\V_{\a\b,MN}^{rs} a_N^{(s)\b\dagger}\label{E'}
\ee
We have introduced the indices $M=\{0,m\}, N=\{0,n\}$ and the
vacuum $|\tilde 0\rangle = 
|0\rangle \otimes |\Omega_{b,\theta}\rangle$, where 
$| \Omega_{b,\theta}\rangle$ is the vacuum with respect to the 
oscillators
\be
a_0^{(r)\alpha} = \frac 12 \sqrt b \hat p^{(r)\alpha} 
- i\frac {1}{\sqrt b} \hat x^{(r)\alpha},
\quad\quad
a_0^{(r)\alpha\dagger} = \frac 12 \sqrt b \hat p^{(r)\alpha} + 
i\frac {1}{\sqrt b}\hat x^{(r)\alpha}, \label{osc}
\ee
where $\hat p^{(r)\alpha}, \hat x^{(r)\alpha}$ are the zero momentum 
and position operator of the $r$--th string; i.e.
$a_0^\alpha|\Omega_{b,\theta}\rangle=0$.
It is understood that $p^{(r)\alpha} = G^{\a\b}p^{(r)}_\b$, and
\be
\big[a_M^{(r)\alpha},a_N^{(s)\beta\dagger}\big]= 
G^{\alpha\beta}\delta_{MN}\delta^{rs} 
\label{aNaN} 
\ee
The coefficients $\V_{MN}^{\a\b,rs}$ are given by
\be
&&\V_{00}^{\a\b,rs} = G^{\a\b}\delta^{rs}- \frac {2A^{-1}b}{4a^2+3}
\left(G^{\a\b} \phi^{rs} -ia \epsilon^{\a\b}\chi^{rs}\right)
\label{VV00}\\
&&\V_{0n}^{\a\b,rs} = \frac {2A^{-1}\sqrt b}{4a^2+3}\sum_{t=1}^3
\left(G^{\a\b} \phi^{rt} -ia\epsilon^{\a\b}\chi^{rt}\right)
V_{0n}^{ts}\label{VV0n}\\
&&\V_{mn}^{\a\b,rs} = G^{\a\b}V_{mn}^{rs}-
\frac {2A^{-1}}{4a^2+3}\sum_{t,v=1}^3
V_{m0}^{rv}\left(G^{\a\b} \phi^{vt} 
-ia \epsilon^{\a\b}\chi^{vt}\right)V_{0n}^{ts}\label{VVmn}
\ee
Here, by definition, $V_{0n}^{rs}=V_{n0}^{sr}$, and 
\be
\phi^{rs}= \left(\matrix{1& -1/2& -1/2\cr 
                    -1/2& 1& -1/2\cr
                     -1/2 &-1/2 &1}\right),\quad\quad
\chi^{rs}= \left(\matrix{0&1&-1\cr -1&0&1\cr 1&-1&0}\right)
\label{phichi}
\ee
These two matrices satisfy the algebra
\be
\chi^2 = - 2\phi,\quad\quad \phi\chi=\chi\phi = \frac 32 \chi,\quad\quad 
\phi^2= \frac 32 \phi\label{chiphi}
\ee
Moreover, in (\ref{VVmn}), we have introduced the notation
\be 
A = V_{00}+ \frac b2,  \quad\quad\quad
a = -\frac {\pi^2}A\, B,
\label{definitions}
\ee

Next we introduce the twist matrix $C'$ by $C'_{MN}= (-1)^M
\delta_{MN}$ and define
\be
\X^{rs} \equiv C' \V^{rs}, \quad\quad r,s=1,2,
\quad\quad \X^{11}\equiv \X\label{X}
\ee
These matrices commute
\be
[\X^{rs}, \X^{r's'}] =0 \label{commute}
\ee
Moreover we have the following properties, which mark a difference
with the $B=0$ case,
\be
C'\V^{rs}= \tilde \V^{sr}C' ,\quad\quad C'\X^{rs}= \tilde \X^{sr}C'
\ee
where tilde denotes transposition with respect to the $\a,\b$ 
indices. Finally one can prove that
\be
&&\X+ \X^{12}+ \X^{21} = \I\0\\
&& \X^{12}\X^{21} = \X^2-\X\0\\
&& (\X^{12})^2+ (\X^{21})^2= \I- \X^2\0\\
&& (\X^{12})^3+ (\X^{21})^3 = 2 \X^3 -  3\X^2 +\I\label{Xpower}
\ee
In the matrix products of these identities, as well as throughout
the paper, the indices $\a,\b$ must be understood in alternating
up/down position: $\X^{\a}{}_{\b}$. For instance, in (\ref{Xpower})
$\I$ stands for $\delta^\a{}_\b\,\delta_{MN}$.

The lump solution we found in \cite{BMS} satisfies 
$|S\rangle=|S\rangle * |S\rangle$ and can be written as
\be
|S\rangle \!&=&\! \left\{{\rm Det}(1-X)^{1/2}{\rm Det} (1+T)^{1/2}\right\}^{24}
{\rm exp}\left(-\frac 12 \eta_{\bar \mu\bar \nu}\sum_{m,n\geq 1} 
a_m^{\bar \mu\dagger}S_{mn}a_n^{\bar \nu\dagger}\right)|0\rangle
\otimes\label{fullsol}\\
&& \frac {A^2 (3+4a^2)}{\sqrt{2 \pi b^3}({\rm Det}G)^{1/4}} 
\left( {\rm Det}(\I -\X)^{1/2}{\rm Det}(\I + \T)^{1/2}\right)
{\rm exp}\left(-\frac 12 \sum_{M,N\geq 0}
a_M^{\a\dagger}\ES_{\a\b,MN}a_N^{\b\dagger}\right)
|\tilde 0 \rangle,\0
\ee
The quantities in the first line are defined in ref.\cite{RSZ2} 
with $\bar\mu,\bar\nu=0,\ldots, 23$ denoting the parallel 
directions to the lump and the matrix $\ES= C'\T $ is given by
\be
\T = \frac 1{2\X}\left( \I +\X - \sqrt{(\I + 3\X)(\I-\X)}
\right) \label{sol2}
\ee
This is a solution to the equation
\be
\X \T^2 - (\I + \X)\T + \X=0\label{equation}
\ee

The solution (\ref{fullsol}) is interpreted as a D--23 brane.
Let $\mathfrak e$ denote the energy per unit volume, 
which coincides with the brane tension when $B=0$. Then one can 
compute the ratio of the D23--brane energy density 
${\mathfrak e}_{23}$ to the D25-brane 
energy density ${\mathfrak e}_{25}$ (from \cite{RSZ1}):
\be
\frac {{\mathfrak e}_{23}}{ {\mathfrak e}_{25}} &\!=\!&
\frac {(2\pi)^2} {({\rm Det}G)^{1/4}}\cdot 
{\EuScript{R}} \label {ratio1} \\
\R &\!=\!&  
\frac {A^4 (3+4a^2)^2}{2 \pi b^3({\rm Det}G)^{1/4}}
\frac {{\rm Det}(\I -\X)^{3/4}{\rm Det}(\I + 3\X)^{1/4}}
{{\rm Det}(1 -X)^{3/2}{\rm Det}(1 + 3X)^{1/2}}\label{ratio2}
\ee
where $X\equiv X^{11}=CV^{11}$ is the matrix called $M$ in \cite{RSZ2}. 
$X$ is the matrix relevant to the sliver solution in VSFT. The 1 
in the denominator of (\ref{ratio2}) stands for $\delta_{nm}$.

It was conjectured in \cite{BMS} that $\R$ equals 1 so that
the ratio (\ref{ratio1}) is 
exactly what is expected for the ratio of a flat static D25--brane
energy density and a D23--brane energy density in the 
presence of the $B$ field (\ref{B}). This is indeed so as we 
will prove in the next section. 

\section{Proof that $\R=1$}

This section is devoted to the proof of
\be
\R =1\label{R=1}
\ee
What we need is compute the ratio of ${\rm Det} (\I -\X)$ and
${\rm Det} (\I +3\X)$ with respect to the squares of 
${\rm Det} (1 -X)$ and ${\rm Det} (1 +3X)$, respectively.  
To this end we follow the lines of ref.\cite{Oku3}.  
To start with we rewrite $\V^{11}\equiv \V$ in a more convenient 
form. Following \cite{Oku3}, we introduce the vector notation
$|v_e\rangle$ and $|v_0\rangle$ by means of
\be
|v_e\rangle_n = \frac {1+(-1)^n}2 \frac {A_n}{\sqrt n},
\quad\quad
|v_o\rangle_n = \frac {1-(-1)^n}2 \frac {A_n}{\sqrt n},\0
\ee
The constants $A_n$ are as in \cite{GJ1}. Now we can write
\be
&&\V_{00} = \left(1- \frac {2 A^{-1}b}{4a^2+3}\right)\, {\bf 1}
\0\\
&&\V_{0n} = - \frac{2 A^{-1}\sqrt{2b}}{4a^2+3}\,{\bf 1}\,
\langle v_e|_n
+ i {\sqrt \frac{2b}3} \frac {4a A^{-1}}{4a^2+3}\, {\bf e}\,
\langle v_o|_n,\quad\quad \V_{0n}= (-1)^n \V_{n0}\label{V}\\
&& \V_{nm}= \left(V_{nm} - \frac{4 A^{-1}}{4a^2+3}
(|v_e\rangle \langle v_e| +|v_o\rangle\langle v_o|)_{nm}\right)
\,{\bf 1}\, + i \frac 8{\sqrt 3} \frac {a A^{-1}}{4a^2+3}
(|v_e\rangle\langle v_o| -|v_o\rangle\langle v_e|)_{nm}
\,{\bf e}\0
\ee
where we have understood the indices $\a,\b$. They can be reinserted
using
\be
{\bf 1}^\a{}_\b = \delta^\a{}_\b, \quad\quad {\bf e}^\a{}_\b = 
\epsilon^\a{}_\b \0
\ee
Now $\X = C'\V$ can be written in the following block matrix form
\be
\X= \left(\matrix{ (1-2 Kb){\bf 1} && 
{-2K\sqrt {2b} \,\one\, \langle v_e| 
+ 4 ia K  \sqrt{\frac {2b}3}\, \e\, \langle v_o|}\cr
{}&{}&\cr
\matrix{-2K\sqrt {2b} |v_e\rangle\, \one\cr
 + 4i a K  \sqrt{\frac {2b}3}|v_o\rangle\, \e \cr}
&&\matrix {X \one - 4K\, \one\,
(|v_e\rangle \langle v_e| -|v_o\rangle\langle v_o|)\cr
+ \frac 8{\sqrt 3} ia K\, \e\, (|v_e\rangle\langle v_o| +
|v_o\rangle\langle v_e|)\cr}\cr}\right)
\label{bigX}
\ee
where all $m,n$ as well as all $\a,\b$ indices are understood, 
$K=\frac {A^{-1}}{4a^2+3}$.

The first determinant we have to compute is the one of the matrix
$\I-\X$. Using (\ref{bigX}) we extract from $\I-\X$ the factor
$2bK$ and represent the rest in the block form
\be
\frac 1{2bK} (\I-\X) = \left(\matrix {{\cal A} &{\cal B}\cr 
{\cal C}&{\cal D}\cr}\right)\0
\ee
By a standard formula, the determinant of the RHS is given by the determinant
of ${\cal D}-{\cal C}{\cal A}^{-1}{\cal B}$. After some algebra and 
using the obvious identity $\langle v_o|v_e\rangle=0$, one gets
\be
{\cal D}-{\cal C}{\cal A}^{-1}{\cal B}\!&=&\! 
\left(\matrix{  1-X  -\frac 43 A^{-1}|v_o\rangle\langle v_o| & 0\cr
0&1-X -\frac 43 A^{-1}|v_o\rangle\langle v_o| \cr} \right)\0\\
\!&=&\! \left(1-X -\frac 43 A^{-1}|v_o\rangle\langle v_o|\right)\, 
\one\0
\ee
The rest of the computation is straightforward,
\be
{\rm Det}(\I-\X) \!&=&\! (2bK)^2 \left( {\rm Det}\Big(1-X - \frac 43 A^{-1}
|v_o\rangle\langle v_o|\Big)\right)^2\0\\
\!&=&\! (2bK)^2 \left({\rm Det}(1-X)\right)^2 
\left({\rm Det}\Big(1 - \frac 43 A^{-1}\, \frac 1{1-X}\,
|v_o\rangle\langle v_o|\Big)\right)^2 \0\\
\!&=&\! \left(\frac bA \right)^4
\left(\frac{1}{4a^2+3}\right)^2
 \left({\rm Det}(1-X)\right)^2 
\label{det1-X}
\ee
In the last step we have used the identities, see \cite{Oku3},
\be
{\rm Det}\left(1 - \frac 43 A^{-1}\, \frac 1{1-X}\,
|v_o\rangle\langle v_o|\right)= 1 - \frac 43 A^{-1}\, 
\langle v_o|\frac 1{1-X}\,|v_o\rangle\label{det}
\ee
and
\be
\langle v_o|\frac 1{1-X}\,|v_o\rangle = \frac 34 V_{00}\label{V00}
\ee

The treatment of ${\rm Det} (\I +3\X)$ is less trivial.
We start again by writing $ (\I +3\X)$ in block matrix form
\be
\I +3\X= \left(\matrix{ (4-6 Kb){\bf 1} && 
{-6K\sqrt {2b} \,\one\, \langle v_e|
+ 4 ia K  \sqrt{6b}\, \e\, \langle v_o| }\cr
{}&{}&\cr
\matrix{-6K\sqrt {2b}\, |v_e\rangle\, \one\cr
 + 4i a K  \sqrt{6b}\,|v_o\rangle\, \e \cr}
&&\matrix {(1+3X) \one - 12K\, \one\,
(|v_e\rangle \langle v_e| -|v_o\rangle\langle v_o|)\cr
+ 8{\sqrt 3} ia K\, \e\, (|v_e\rangle\langle v_o| +
|v_o\rangle\langle v_e|)\cr}\cr}\right) 
\label{big3X}
\ee
and set
\be
\I +3\X\equiv  (4-6bK)\left(\matrix {{\cal A} &{\cal B}\cr 
{\cal C}&{\cal D}\cr}\right) 
\ee
Therefore
\be
{\rm Det}(\I +3\X)\!&=&\! (4-6bK)^2 \det\left({\cal D}-
{\cal C}{\cal A}^{-1}{\cal B} \right)\0\\
&=&\!(4-6bK)^2\,
\left(\det (1+3X)\right)^2\,\det \left( \frac 1{1+3X}
\left({\cal D}-{\cal C}{\cal A}^{-1}{\cal B} \right)\right)
\label{eq1}
\ee
The last expression is formal. In fact $X$ has an eigenvalue $-\frac 13$
which renders the RHS of (\ref{eq1}) ill--defined. To avoid this
we follow \cite{Oku3} and introduce the regularized inverse
\be
Y_\varepsilon = \frac 1{1+3X- \varepsilon^2 X}\label{Y}
\ee
where $\varepsilon$ is a small parameter, and replace it into 
(\ref{eq1}). After some algebra we find
\be
Y_\varepsilon \,\left({\cal D}-{\cal C}{\cal A}^{-1}{\cal B} \right)=
\A\cdot \B\label{eq2}
\ee
The matrices in the RHS are given by
\be
\A = \left(\matrix{1+ \a \Y |v_e\rangle \langle v_e| +
\b\Y |v_o\rangle \langle v_o| &0\cr
0 & 1+ \a \Y |v_e\rangle \langle v_e| +\b\Y |v_o\rangle \langle v_o| 
\cr}\right)\label{A}
\ee
where
\be
\a = - \frac {24 K}{2-3bK},\quad\quad 
\b = 12 K \frac {2- A^{-1}}{2-3bK},
\label{ab}
\ee 
and
\be
\B = \left(\matrix{1& \lambda\Y |v_2\rangle \langle v_o| 
+ \mu\Y |v_o\rangle \langle v_e|\cr
 -\lambda\Y |v_e\rangle \langle v_o| 
- \mu\Y |v_o\rangle \langle v_e|& 1\cr}\right)\label{Bscript}
\ee
where,
\be
\lambda= \frac {\gamma}{1 +\a \langle v_e|\Y|v_e\rangle},
\quad\quad \mu = \frac {\gamma}{1 +\b \langle v_0|\Y|v_0\rangle},
\quad\quad \gamma^2 + \a\b = -\frac 4{V_{00}} \b\label{lm}
\ee
Now, after some computation,
\be
\det \A = \left(1 +\a \langle v_e|\Y|v_e\rangle\right)^2 
\, \left(1 +\b \langle v_o|\Y|v_o\rangle\right)^2 \label{A2}
\ee
and
\be
\det \B = \left( 1+ \frac {\gamma^2 \langle v_e|\Y|v_e\rangle
\langle v_o|\Y|v_0\rangle}{\left(1 +\a \langle v_e|\Y|v_e\rangle\right)
\, \left(1 +\b \langle v_o|\Y|v_o\rangle\right)}\right)^2\label{B2}
\ee
As a consequence
\be
\det \A\, \det \B = \left( 1 + \a \langle v_e|\Y|v_e\rangle+
\b \langle v_o|\Y|v_0\rangle  \left( 1- \frac 4{V_{00}}
\langle v_e|\Y|v_e\rangle\right)\right)^2\label{eq3}
\ee
Now we can remove the regulator $\varepsilon$ by using the basic 
result of \cite{Oku3}:
\be
\lim_{\varepsilon \rightarrow 0}
\left( 1- \frac 4{V_{00}}\langle v_e|\Y|v_e\rangle\right)
\,\langle v_o|\Y|v_o\rangle = \frac {\pi^2}{12 V_{00}}
\label{oku}
\ee
and
\be
\langle v_e|\frac 1{1+3X}|v_e\rangle =\frac {V_{00}}4.\0
\ee
Inserting this result in (\ref{eq3}) we find
\be
\det \A\, \det \B = \frac {A^2}{(8a^2A+6V_{00})^2}\left(8a^2 + 
\frac{2\pi^2}{A^2}\right)^2
\label{eq4}
\ee
As a consequence of eqs.(\ref{eq1},\ref{eq2},\ref{eq3},\ref{eq4})
we find
\be
\frac{{\rm Det}(\I +3\X)}{({\rm Det}(1 +3X))^2} =
\frac 4{(4a^2+3)^2} \left(8a^2 + 
\frac{2\pi^2}{A^2}\right)^2\label{eq5}
\ee
Finally, substituting this and (\ref{det1-X}) into $\R$, we get
\be
\R &\!=\!&  
\frac {A^4 (3+4a^2)^2}{2 \pi b^3({\rm Det}G)^{1/4}}
\frac {{\rm Det}(\I -\X)^{3/4}{\rm Det}(\I + 3\X)^{1/4}}
{{\rm det}(1 -X)^{3/2}{\rm Det}(1 + 3X)^{1/2}}=1\label{final}
\ee

This is what we wanted to show. It implies
\be
\frac {{\mathfrak e}_{23}}{ {\mathfrak e}_{25}} &\!=\!&
\frac {(2\pi)^2} {({\rm Det}G)^{1/4}}\label{ratio3}
\ee   
which corresponds to the expected result for this ratio, as explained 
in \cite{BMS}. We remark that (\ref{eq5}) implies that the eigenvalue
$-\frac 13$ is also contained in the spectrum of $\X$ with double
multiplicity with respect to $X$.

\section{Some results in VSFT with $B$ field}

In this section we deal with a couple of results which are natural
extensions of analogous results with $B=0$.

\subsection{Wedge--like states}

Wedge states were introduced in \cite{FO}. They are geometrical
states in that they can be defined simply by means of a conformal
map of the unit disk to a portion of it. They are spanned by an integer
$n$: the limit for $n\to \infty$ is the sliver $\Xi$, which is interpreted
as the D25--brane. Wedge states also admit a representation
in terms of oscillators $a_n^\dagger$ with $n>0$,
\be
|W_n\rangle = {\cal N}_n^{26} e^{-\frac 12 a^\dagger CT_na^\dagger}
|0\rangle\label{wn}
\ee
which is specified by the matrix $T_n$, $n>1$. It can be shown that, see
\cite{FO}, $T_n$ satisfy a recursive relation which can be solved
in terms of the matrix $T$ characterizing the sliver state ($T=CS$,
$S$ being the sliver matrix). The normalization 
${\cal N}$
can also be derived from a recursion relation. Since all these results are
essentially based on equations which can be generalized to the case when
a $B$--field is present and were in fact derived in \cite{BMS}, it is easy to 
deduce that analogous results will hold also when a $B$ field is turned on.

The generalized wedge states will be the tensor product of a factor like 
(\ref{wn}) for the the 24 directions in which the components of the $B$ field
are zero and 
\be
|\W_n\rangle = \N_n^2\, e^{-\frac 12 a^\dagger C'\T_na^\dagger}
|\tilde 0\rangle\label{Bwn}
\ee
for the other two directions. From now on we will be concerned 
with the determination of $\T_n$ and $\N_n$. 
We start from the hypothesis that
\be
[\X^{rs},\T_n]=0,\quad\quad C'\T_n = \tilde \T_n C' \label{hypo}
\ee
whose consistency we will verify a posteriori.
 
Now we define $\T_2=0$ and the sequence of states
\be
|\W_{n+1}\rangle = |\W_n\rangle * |\W_2\rangle\label{rec}
\ee
Using eq.(4.4) and (4.7) of \cite{BMS}, with $\Sigma = 
\left(\matrix{C'\tilde \T_n &0 \cr 0&0\cr}\right)$, we find the 
recursion relation
\be
\T_{n+1} \!&=&\! \X^{11}+ (\X^{12},\X^{21}) \left( 1 - 
\left(\matrix{\T_n \X^{11}& \T_n \X^{12}\cr 0&0\cr}\right)\right)^{-1}
\left(\matrix{\T_n \X^{21}\cr 0\cr}\right)\0\\
\!&=&\! \X \, \frac{1- \T_n}{1-\T_n \X}\label{recrel}
\ee
where use has been made of the second equation in (\ref{Xpower}).
Solving this recursion relation, \cite{FO}, we can write
\be
\T_n = \frac {\T + (-\T)^{n-1}}{1- (-\T)^n}\label{Tn}
\ee
Notice that this sequence of states can be extended to 
$|\W_1 \rangle$ defined
by $\T_1 =1$. An analogous recursion relation applies also to 
the normalization factors.
Once solved, it gives
\be
\N_n = K_2^{-1} \det \left( \frac {1-\T^2}
{1-(-\T)^{n+1}}\right)^{1/2}\label{norm}
\ee
The constant $K_2$ is defined in eq.(2.19) of \cite{BMS}.
The relations (\ref{hypo}) are now easy to verify.

The limit of $\T_n$ as $n\to \infty$ is $\T$ (i.e. the deformation 
of the lump),
provided $\lim\,  \T^n =0$. In turn, the latter holds if the 
eigenvalues of $\T$ are in absolute value less
then 1, as those of $T$ are. This is very likely in view of 
the results on the ratio of 
determinants in the last section \footnote{An analysis of the 
eigenvalues of $\X$ has 
been recently announced in \cite{Bofeng}.}.

\subsection{Orthogonal projectors with $B$ field}
 
In the presence of a background $B$ field it is also possible to
construct other projectors than the one shown in (\ref{fullsol}).
To show this we follow ref.\cite{RSZ3}. The treatment is 
very close to sec.3 and 5 of that reference, and the main purpose
of this subsection is to stress some differences with it.   
As usual we will be concerned only with the transverse part of the
projectors, the parallel being exactly the same as in \cite{RSZ3},
and will denote the transverse part of the solution (\ref{fullsol})
by $|\ES_\perp\rangle$. 

We start by introducing the projection operators
\be
\rho_1 \!&=&\! \frac 1{(\I +\T)(\I-\X)} \left[ \X^{12} (\I-\T\X)
+\T (\X^{21})^2\right]\label{rho1}\\
\rho_2 \!&=&\! \frac 1{(\I +\T)(\I-\X)} \left[ \X^{21} (\I-\T\X)
+\T (\X^{12})^2\right]\label{rho2}
\ee
They satisfy
\be
\rho_1^2 = \rho_1,\quad\quad \rho_2^2 = \rho_2, \quad\quad 
\rho_1+\rho_2 = \I\label{proj}
\ee
i.e. they project onto orthogonal subspaces. Moreover, if we
use the superscript $^T$ to denote transposition with respect to
the indices $N,M$ and $\a,\b$, we have
\be
\rho_1^T=\tilde\rho_1 = C'\rho_2 C',\quad\quad 
 \rho_2^T=\tilde\rho_2 = C'\rho_1 C'.\label{rhorels}
\ee

Now, in order to find another solution of the equation 
$|\Psi\rangle * |\Psi\rangle= |\Psi\rangle$, distinct from 
$|\ES_\perp\rangle$, we make the
following ansatz:
\be
|\EP_\perp\rangle = \left(- \xi\tau a^\dagger \,\, \zeta \cdot
a^\dagger + \kappa\right) |\ES_\perp\rangle\label{ansatz}
\ee
where $\xi = \{\xi_N^\a\}$, $\zeta= C'\xi$ and $\tau$ is the matrix
$\left(\matrix{1&0\cr 0& -1\cr}\right)$ acting on the indices $\a$ and $\b$. $\kappa$ is a constant
to be determined and $\xi$ is required to satisfy the constraints:
\be
\rho_1 \xi =0,\quad\quad \rho_2 \xi =\xi, \quad\quad
{\rm i.e.}\quad \tilde \rho_1 \zeta =\zeta, 
\quad\quad \tilde \rho_2 \zeta=0\label{rhozeta}
\ee
Using (\ref{proj},\ref{rhozeta}) it is simple to prove that
\be
\zeta^T \, f(\X^{rs}, \T)\, \xi =0 , \quad\quad
\xi^T \, f(\tilde \X^{rs}, \tilde \T) \zeta =0\0
\ee
for any function $f$. Now, imposing $|\EP_\perp\rangle * 
|\ES_\perp\rangle =0$ we determine $\kappa$:
\be
\kappa = -\frac 12 \zeta^T\tau
\left({\cal V} {\cal K}^{-1}\right)_{11}
\xi - \frac 12 \xi^T\left({\cal V} {\cal K}^{-1}\right)_{11}
\tau\zeta\label{kappa}
\ee
where
\be
{\cal K} = \I -\T\X,\quad\quad
{\cal V} = \left(\matrix{{\V}^{11}& {\V}^{12}\cr
{\V}^{21}& {\V}^{22}\cr}\right)\label{KV}
\ee 
Next we compute $|\EP_\perp\rangle * |\EP_\perp\rangle $.
This gives
\be
|\EP_\perp\rangle * |\EP_\perp\rangle =
\frac 12 \left(\xi^T \left({\cal V}{\cal K}^{-1}\right)_{12}\tau \zeta 
+  \zeta^T\tau \left({\cal V}{\cal K}^{-1}\right)_{21} \xi \right)
\left(- a^\dagger\tau\xi \,\, a^\dagger\cdot \zeta + \kappa\right)
|\ES_\perp\rangle \label{pstarp}
\ee
where use has been made of the identities
\be
\zeta^T \tau\left({\cal V}{\cal K}^{-1}\right)_{11} \xi \!&=&\!
\zeta^T \tau\left({\cal V}{\cal K}^{-1}\right)_{22} \xi =
- \zeta^T \tau\left({\cal V}{\cal K}^{-1}\right)_{12} \xi =
\xi^T \tau\frac {\T}{\I-\T^2}\xi\0\\
\xi^T \left({\cal V}{\cal K}^{-1}\right)_{11}\tau \zeta \!&=&\!
\xi^T \left({\cal V}{\cal K}^{-1}\right)_{22}\tau \zeta =
- \xi^T \left({\cal V}{\cal K}^{-1}\right)_{21}\tau \zeta =
\zeta^T \frac {\T}{\I-\T^2}\tau\zeta\label{ids}\\
\xi^T \left({\cal V}{\cal K}^{-1}\right)_{12} \tau\zeta \!&=&\!\zeta^T
\frac 1{\I-\T^2} \tau\zeta, \quad\quad
\zeta^T \tau\left({\cal V}{\cal K}^{-1}\right)_{21} \xi =
\xi^T \tau\frac 1{\I-\T^2}\xi\0
\ee
Similarly one can prove that
\be
\zeta^T \frac {\T}{\I-\T^2}\tau\zeta= 
\xi^T \tau\frac {\T}{\I-\T^2}\xi,\quad\quad
\zeta^T \frac 1{\I - \T^2} \tau\zeta =\xi^T \tau
\frac 1{\I - \T^2} \xi\label{agg}
\ee
So, in order for $|\EP_\perp\rangle$ to be a projector,
we have to impose
\be
\left(\xi^T \left({\cal V}{\cal K}^{-1}\right)_{12}\tau \zeta 
+  \zeta^T \tau\left({\cal V}{\cal K}^{-1}\right)_{21} \xi \right)=
2\,\xi^T \tau \frac 1{\I - \T^2} \xi=2
\label{cond1}
\ee
 
Using this and following \cite{RSZ3}, it is simple to prove that
\be
\langle \EP_\perp | \EP_\perp\rangle = 
\left(\zeta^T \frac 1{\I - \T^2} \tau\zeta \,\, 
\xi^T \tau\frac 1{\I - \T^2} \xi\right)
\,\langle \ES_\perp | \ES_\perp\rangle = \,
\langle \ES_\perp | \ES_\perp\rangle\label{normeq}
\ee
thanks to (\ref{agg},\ref{cond1}).

Therefore, under the condition
\be
\xi^T \tau\frac 1{\I - \T^2} \xi=1\label{onlycond}
\ee
the BPZ norm of 
$|\EP_\perp\rangle + |\ES_\perp\rangle$ is twice the norm 
of $|\ES_\perp\rangle$.
As a consequence the sum of these two states, once they are tensored
by the corresponding 24--dimensional complements defined in
\cite{RSZ3}, represent a couple of parallel D23-branes.

Similarly one can construct more complicated brane configurations
as suggested in \cite{RSZ3}.

\section{Some effects of the $B$ field}

So far we have seen that many results of VSFT without $B$ field
are accompanied by parallel results in the presence of the $B$ field.
In this section we would like to show that this is not merely a 
formal replication, but that in some cases a nonvanishing background 
$B$ field affects in a significant way the form of such results. 
Precisely we would like to show that a $B$ field has the effect of
smoothing out some of the singularities that appear in the VSFT.  

\subsection{Low energy limit}

In \cite{MT} it was shown that the geometry  of the 
lower--dimensional lump states representing Dp-branes
is singular. This can be seen both in the zero slope limit 
$\alpha' \rightarrow 0$ and as an exact result. It can be 
briefly stated by saying that the midpoint of the string is
confined on the hyperplane of vanishing transverse
coordinates. It is therefore interesting to see whether
the presence of a $B$ field modifies this
situation. Moreover it is also well known that soliton solutions of field theories defined on a noncommutative space describe Dp-branes (\cite{GMS}, \cite{Komaba}). It is then interesting
to see if we can recover at least the simplest GMS soliton, using the particular low energy limit, i.e. the  limit of \cite{SW}, that gives a noncommutative field theory from a string theory with a $B$ field turned on.

To discuss this limit we first reintroduce the closed string metric $g_{\a\b}$ as $ g\,\delta_{\a\b}$. Now we take $\alpha' B \gg g$, in such a way that $G$, $\theta$ and $B$ are kept fixed. The limit is described by means of a parameter $\epsilon$ going to $0$ as in \cite{MT} ($\alpha' \sim \epsilon$). 
 We could also choose to parametrize the $\alpha' B \gg g$ condition by sending $B$ to infinity, keeping $g$ and $\alpha'$ fixed and operating a rescaling of the string modes as in \cite{Sch}, of course at the end we get identical results. 
By looking at the exponential of the 3-string field theory vertex in the presence of a $B$ field
\be
&&\sum_{r,s=1}^3\left(\frac 12 \sum_{m,n\geq 1}
G_{\alpha\beta}a_m^{(r)\alpha\dagger}V_{mn}^{rs}
a_n^{(s)\beta\dagger}  
+ \sqrt{\alpha'}\,\sum_{n\geq 1}G_{\alpha\beta}p_{(r)}^{\alpha}V_{0n}^{rs}
a_n^{(s)\beta\dagger} \0 \right. \\
&& \left. \quad \quad  + \,\alpha'\,\frac 12 G_{\alpha\beta}p_{(r)}^{\alpha}V_{00}^{rs}
p_{(s)}^\beta + 
\frac i2 \sum_{r<s} p_\alpha^{(r)}\theta^{\alpha\beta}
p_\beta^{(s)}\right)
\ee
we see that the limit is characterized by the rescalings 
\be
&& V_{mn}  \rightarrow  V_{mn} \0  \\
&& V_{m0}  \rightarrow  \sqrt{\epsilon} V_{m0}  \\
&& V_{00}  \rightarrow  \epsilon V_{00} \0
\ee
$G_{\a\b}$ and $\theta^{\a\b}$  are kept fixed. Their explicit dependence on $g$, $\alpha'$ and $B$ will be reintroduced at the end of our calculations in the form
\be
G_{\a\b}  = \frac{(2\pi\alpha'B)^2}{g}\delta_{\a\b}, \quad\quad
 \theta   =  \frac{1}{B} \label{SWGB}
\ee  
Substituting the leading behaviors of $V_{MN}$ in eqs.(\ref{VVmn}), 
and keeping in mind that $A = V_{00} + \frac{b}{2}$, the coefficients 
$\V_{MN}^{\a\b,rs}$ become
\be
\V_{00}^{\a\b,rs} & \rightarrow &\, G^{\a\b}\delta^{rs}- \frac {4}{4a^2+3}
\left(G^{\a\b} \phi^{rs} -ia \hat\epsilon^{\a\b}\chi^{rs}\right)\\
\V_{0n}^{\a\b,rs} & \rightarrow &\, 0\\
\V_{mn}^{\a\b,rs} & \rightarrow &\, G^{\a\b}V_{mn}^{rs}\label{VVmnt}
\ee
We see that the squeezed state (\ref{fullsol}) factorizes in two parts: 
the coefficients $\V_{mn}^{\a\b,11}$ reconstruct the full 25 
dimensional sliver, while the coefficients $\V_{00}^{\a\b,11}$ 
take a very simple form  
\be
&&\ES_{00}^{\a\b}=\frac{2 |a| -1}{2 |a| +1}\, G^{\a\b} \equiv s\, 
G^{\a\b} \label{expB}
\ee
The soliton lump with this choice of the coefficients $\V_{MN}^{\a\b,rs}$ 
will be called $|\hat{\ES}\rangle$ 
\be
|\hat{\ES}\rangle \!&=&\! \left\{\det(1-X)^{1/2}\det (1+T)^{1/2}\right\}^{24}
{\rm exp}\left(-\frac 12 \eta_{\bar \mu\bar \nu}\sum_{m,n\geq 1} 
a_m^{\bar \mu\dagger}S_{mn}a_n^{\bar \nu\dagger}\right)|0\rangle
\otimes\label{fullsollow}\\
&& {\rm exp}\left(-\frac 12 G_{\a \b}\sum_{m,n\geq 1} 
a_m^{\a \dagger}S_{mn}a_n^{\b \dagger}\right)|0\rangle \otimes \0\\
&& \frac {A^2 (3+4a^2)}{\sqrt{2 \pi b^3}({\rm Det}G)^{1/4}} 
\left( {\rm Det}(\I -\X)^{1/2}{\rm Det}(\I + \T)^{1/2}\right)
{\rm exp}\left(-\frac 12 
s 
a_0^{\a\dagger}G_{\a\b}a_0^{\b\dagger}\right)| \Omega_{b, \theta}\rangle,\0
\ee
where $\bar \mu, \bar\nu = 0, \dots 23$ and $\a, \b = 24, 25$. 
In the low energy limit we have also
\be
&& {\rm Det}(\I -\X)^{1/2}{\rm Det}(\I + \T)^{1/2}= 
 \frac{4}{4a^2 + 3}\, {\rm det} (1-X)\,\,
\frac{4a}{2a+1}\, {\rm det} (1+T) \label{dets}
\ee
So the complete lump state becomes
\be
|\hat{\ES}\rangle \!&=&\! \left\{\det(1-X)^{1/2} \det (1+T)^{1/2}\right\}^{26}
{\rm exp}\left(-\frac 12 G_{ \mu \nu}\sum_{m,n\geq 1} 
a_m^{ \mu\dagger}S_{mn}a_n^{ \nu\dagger}\right)|0\rangle \otimes
\0\\
& & \frac{4a}{2a+1}\,\, \frac{b^2}{\sqrt{2\pi b^3}(\det G)^{1/4}}\,\, 
{\rm exp}\left(-\frac 12 
s 
a_0^{\a\dagger}G_{\a\b}a_0^{\b\dagger}\right)| \Omega_{b, \theta}\rangle,\label{solB}
\ee
where  $\mu, \nu = 0, \dots 25$ and $G_{\mu\nu} = \eta_{\bar\mu \bar\nu}\oplus G_{\a\b}$. The first line of (\ref{solB}) is the usual 25-dimensional sliver up to a simple rescaling of $a_n^{\alpha\dagger}$.  
The norm of the lump is now regularized by the presence of $a$ which is directly proportional to $B$: $a = -\frac{\pi^2}{A}B$. Using  
\be
&& | x \rangle = \sqrt{\frac{2\sqrt{\det G}}{b\pi}} 
\exp\left[-\frac{1}{b}x^{\a}G_{\a\b}x^{\b} 
-\frac{2}{\sqrt{b}}i a_0^{\a\dagger} G_{\a\b}x^{\b}
+\frac{1}{2}a_0^{\a\dagger}G_{\a\b}a_0^{\b\dagger} 
\right]
|\Omega_{b, \theta} \rangle
\ee
we can calculate the projection onto the basis of position 
eigenstates of the transverse part of the lump state
\be
\langle x | e^{-\frac{s}{2}(a_0^{\dagger})^2}|\Omega_{b, \theta} \rangle & = & 
\sqrt{\frac{2\sqrt{\det G}}{b\pi}} \frac{1}{1+s}\,
 e^{-\frac{1-s}{1+s}\frac{1}{b}x^{\a}x^{\b}G_{\a\b}} \0 \\
& = & \sqrt{\frac{2\sqrt{\det G}}{b\pi}} \frac{1}{1+s}\,
      e^{-\frac{1}{2|a|b}x^{\a}x^{\b}G_{\a\b}} \label{xproj}
\ee
The transverse part of the lump state in the $x$ representation is then
\be
\langle x |\hat{\ES}_\perp \rangle  
= \frac{1}{\pi}\, e^{-\frac{1}{2|a|b}x^{\a}x^{\b}G_{\a\b}} \label{regul}
\ee

Using the form (\ref{SWGB}) of $G_{\a\b}$ and $\theta^{\a\b}$  
and the expression for $a$ where we explicitate the dependence on $g$ and $\alpha'$, \cite{BMS}
\be
a = \frac{\theta}{4A}\sqrt{\det G} = - \frac{2 \pi^2 (\alpha')^2 B}{b\, g}
\ee
we obtain the simplest soliton solution
of \cite{GMS} (see also \cite{Komaba} and references therein):
\be
 e^{-\frac{1}{2|a|b}x^{\a}x^{\b}G_{\a\b}} &\rightarrow &
e^{-\frac{x^{\a}x^{\b}\delta_{\a\b}}{|\theta|}} \0
\ee
which corresponds to the $|0\rangle\langle 0|$ projector in 
the harmonic oscillator Hilbert space of \cite{GMS}.
We notice that the profile and the normalization of 
$\langle x |\hat{\ES}_\perp \rangle$ do not depend on $b$. 

As compared to \cite{MT}, the $B$ field provides a natural realization of the regulator for the tachyonic soliton introduced ad hoc there. This beneficial effect of the $B$ field  
is confirmed by the fact that the projector (\ref{solB}) is no longer 
annihilated by $x_0$
\be 
x_{0}\,\, {\rm exp}\left(-\frac 12 
s a_0^{\a\dagger}G_{\a\b}a_0^{\b\dagger}\right)| \Omega_{b, \theta}\rangle
& = & i\,\frac{\sqrt{b}}{2}\,(a_0 - a_0^{\dagger})\,
{\rm exp}\left(-\frac 12 
s a_0^{\a\dagger}G_{\a\b}a_0^{\b\dagger}\right)| \Omega_{b, \theta}\rangle \0 \\
& =& -
i\,\frac{\sqrt{b}}{2}\,
\left[\frac{4a}{2a+1}\right]a_0^{\dagger}\, {\rm exp}\left(-\frac 12 
s a_0^{\a\dagger}G_{\a\b}a_0^{\b\dagger}\right)| \Omega_{b, \theta}\rangle \0
\ee
Therefore, at least in the low energy limit, the singular structure 
found in \cite{MT} has disappeared. 

\subsection{The string midpoint}

In the previous subsection a very interesting question has been raised.
It concerns the string midpoint. It was shown in \cite{MT} that,
in the absence of a $B$ field, the string midpoint in the lower 
dimensional lumps is confined to the hyperplane (D--brane) 
of vanishing
transverse coordinates. We have seen above that this is not anymore
the case when a constant background $B$ field is present, at least
in the field theory limit. One might deduce from this that the string
midpoint is not confined anymore in the full VSFT either. However 
such conclusion is far from self--evident. 
As we will see in the sequel, the field theory
limit (tachyon) contribution to the string midpoint is only one out 
of an infinite set of other contributions which characterize the 
full theory and it is not inconceivable that there might be a 
cancellation between the field theory limit contribution and all 
the other terms. Evaluating the exact string midpoint position in 
the full VSFT is in fact a nontrivial and interesting problem. 
We intend to address it in this subsection.

The oscillator expansion for the transverse string coordinates 
is, \cite{sugino}, setting $\a' =\frac 12$,
\be
x^\a(\sigma) = x_0^\a + \frac {\theta^{\a\b}}{\pi}p_{0,\b}
\Big(\sigma - \frac {\pi}2\Big) + \sqrt 2\, \sum_{n=1}^\infty \left[x_n^\a
\,\cos\,(n\sigma) + \frac {\theta^{\a\b}}{\pi}\frac 1n\, p_{n,\b}
\,\sin \,(n\sigma)\right]\label{xsigma}
\ee
Therefore the string midpoint is specified by
\be
x^\a\Big(\frac \pi{2}\Big) = x_0^\a + \sqrt 2 \sum_{n=1}^\infty (-1)^n
\left[x_{2n}^\a - 
 \frac {\theta^{\a\b}}{\pi}\frac 1{2n-1}\, p_{2n-1,\b}
\right]\label{midpoint}
\ee
It is more convenient to pass to the operator basis 
$a_N^\a,a_N^{\a\dagger}$, which satisfies the algebra (\ref{aNaN})
and are related to $x_n, p_n$ by
\be
x_n^\a = \frac i{\sqrt{2n}}\left( a_n^\a - a_n^{\a\dagger}\right),
\quad\quad 
p_{n,\a}= \sqrt \frac n2 G_{\a\b} 
\left( a_n^\b + a_n^{\b\dagger}\right),\label{xpaa}
\ee
while the analogous relation for $x_0,p_0$ is given by eq.(\ref{osc})
with the specification that throughout this section, for
simplicity, we fix $b=2$.

Now, confinement of the string midpoint means
\be
x^\a\Big(\frac {\pi}2\Big)\,|\ES_\perp\rangle = 0\label{confin}
\ee
Evaluating the LHS we get
\be
x^\a\Big(\frac {\pi}2\Big)\,|\ES_\perp\rangle \!&=&\! - \frac i{\sqrt 2}
(a^\dagger + a^\dagger \ES )_0^\a |\ES_\perp\rangle 
- i \sum _{n=1}^\infty \frac {(-1)^n}{\sqrt {2n}} 
(a^\dagger + a^\dagger \ES )_{2n}^\a |\ES_\perp\rangle \0\\
&& - \sum _{n=1}^\infty \frac {(-1)^n}{\sqrt {2n-1}} 
\frac {\theta^{\a\b}}{\pi}G_{\b\gamma} 
(a^\dagger - a^\dagger \ES )_{2n-1}^\gamma |\ES_\perp\rangle 
\label{confin2}
\ee
Confinement requires that this vanish. In order to write this
condition in compact form, we introduce the 
$2\times 2$--matrix--valued vector 
\be
\Theta = |\nu\rangle \, {\bf 1} + |\mu\rangle \,{\bf e}\label{omega}
\ee
where
\be
&&|\nu\rangle = \{\nu_0,\nu_{2n}\}, \quad\quad \nu_0 = \frac 1{\sqrt 2},
\quad\quad \nu_{2n} = \frac {(-1)^n}{\sqrt {2n}} \0\\
&&|\mu\rangle = \{\mu_{2n-1}\},\quad\quad \mu_{2n-1} = i \pi\,B\,
\frac {(-1)^n}{\sqrt{2n-1}} \label{munu}
\ee
Now the confinement condition for the string midpoint can be written
as
\be
\ES C' \,\Theta = - \Theta,\quad \quad{\rm or, \,\,equivalently,}
\quad\quad
\tilde \T \,\Theta= -\Theta,\quad {\rm i.e.}\quad 
\T \,\tilde\Theta=-\tilde \Theta. \label{confin3}
\ee
Due to (\ref{sol2}) an eigenvalue $-1$ of $\T$ corresponds to an
eigenvalue $-\frac 13$ of $\X$ with the same eigenvector.
Let us rewrite $\I+ 3\X$ as
\be
\I +3\X = \EY\, \one + \Z \,{\bf e}\label{1+3x}
\ee
Then eq.(\ref{confin3}) becomes $(\I +3\X)\,\tilde\Theta =0$, which in turn
corresponds to the two equations
\be
&&\EY \,|\nu\rangle + \Z\,|\mu\rangle =0 \label{equ1}\\
&& \Z\,|\nu\rangle - \EY\,|\mu\rangle =0\label{equ2}
\ee
It is useful to further split $\EY$ as $\EY= \EY_0+\EY_1$, where 
$\EY_0=
\EY(B=0)$. Using (\ref{big3X}) one obtains
\be
\EY_0 \,=\, \left( \matrix{ 4(1-A^{-1}) & &- 4 A^{-1} \langle v_e|\cr
&&\cr
- 4 A^{-1} |v_e\rangle  &&1+3X -4\,A^{-1}
(|v_e\rangle\langle v_e| -|v_o\rangle\langle v_o|)\cr}\right)
\label{Y0}
\ee
\be
\EY_1 \,=\, 12\, H\,\left( \matrix{ 1&& \langle v_e|\cr
&&\cr
|v_e\rangle &&|v_e\rangle\langle v_e| -
|v_o\rangle\langle v_o|\cr}\right)
\label{Y1}
\ee
\be
\Z \,=\, 8{\sqrt 3}\,i\,a\,K\, \left( \matrix{ 0&& \langle v_o|\cr
&&\cr
|v_o\rangle  &&|v_e\rangle\langle v_o| +
|v_o\rangle\langle v_e|\cr}\right)
\label{Z}
\ee
where $H= \frac 43 \frac {a^2 A^{-1}}{4a^2+3}$.

Now let us express the previous equations in a more explicit form.
To this end we conform to the notation of section 2 and write
\be
&&|\nu\rangle = \nu_0 \oplus |\nu_e\rangle, \0\\
&&|\mu\rangle = -i \pi B |\lambda_o\rangle, \0
\ee
where
\be
&& |\nu_e\rangle_n = \frac {1+(-1)^n}2 \nu_n, \quad\quad 
\nu_n = \frac {(-1)^{n/2}}{\sqrt n} \label{even}\\
&&|\lambda_o\rangle_n= \frac {1-(-1)^n}2 
\lambda_n,\quad\quad \lambda_n=\frac {(-1)^{(n+1)/2}}{\sqrt n}
\label{odd} 
\ee
We remark that $|\nu\rangle$ is the eigenvector corresponding to
the eigenvalue $-\frac 13$ of $\X(B=0)$, introduced in \cite{MT};
and that $|\lambda_o\rangle$ is the eigenvector with eigenvalue 
$-\frac 13$ of $X$, introduced in \cite{RSZ5}. As a consequence
one has 
\be
\EY_0\,|\nu\rangle =0, \quad\quad \quad (1+3X)|\lambda_o\rangle=0
\label{preveq}
\ee
The first equation can be rewritten as
\be
&&\langle v_e|\nu_e\rangle = V_{00} \nu_0\label{MT1}\\
&&(1+3X)|\nu_e\rangle = 4\nu_0 |v_e\rangle\label{MT2}
\ee
Remarkably enough, all the other equations from (\ref{equ1},
\ref{equ2}), after using (\ref{MT1}) and the second equation in
(\ref{preveq}), reduce to a single one 
\be
\langle v_o|\lambda_o\rangle = {\sqrt \frac 23} \pi\label{fineq}
\ee
Therefore, since eqs.({\ref{preveq}) have been proved independently,
confinement of the string midpoint holds or not according to whether
eq.(\ref{fineq}) is true or not. Now, the LHS of this equation
is
\be
\langle v_o|\lambda_o\rangle = \sum_{n\,\,{\rm odd}} (-1)^{(n+1)/2}
\frac {A_n}n \label{fineq1}
\ee
The latter series can be summed with standard methods and gives
\be
\langle v_o|\lambda_o\rangle = \frac {9-2\sqrt{3} \pi}6\0
\ee
Therefore (\ref{fineq}) is definitely not satisfied. 
So we can conclude that the string midpoint in the presence of
a $B$ field is {\it not confined} on the hyperplane that
identifies the D23--brane.

\section{Conclusion}

In this paper we have shown that the introduction of a $B$ field in
VSFT does not prevent us from obtaining parallel results to those
obtained when $B=0$. Once the formalism is set up, the formal 
complications brought about by the $B$ field are far from scaring.
The calculations of section 3 and 4 are examples of this fact.
On the other hand a nonvanishing background $B$ field may
have advantageous aspects. The smoothing out effects of $B$ on the UV
divergences of noncommutative field theories are well--known.
The aim of this paper was to start exploring the effects of a $B$ 
field on the diverse singularities that appear in VSFT. We have
verified that the singular geometry of the lump solutions,
pointed out in \cite{MT}, disappears in the presence of a $B$
field, in particular the string midpoint is not confined any
longer to stay on the D--brane. 

We remark that this {\it deconfinement} might
mean also that the left--right factorization characteristic of
the sliver solution, \cite{RSZ3,GT,Moeller}, is not possible
for lump solutions with $B$ field. However it looks like
there are other aspects of VSFT which may be fruitfully extended to 
VSFT with $B$ field. For instance, the series of wedge--like states
introduced in section 4.1 seem to suggest that the geometric nature
of the wedge states, \cite{RZ}, persists also in the presence of
a $B$ field. This is confirmed by the results obtained recently in
\cite{MRVY}, where the presence of a $B$ field has been dealt
with entirely geometrically. It would be interesting to know 
whether, for instance, the analogs of butterfly states
in a constant $B$ background, \cite{GRSZ1,GRSZ2,Sch1}, can be 
constructed. Another interesting subject concerns the connection,
if any, of our approach with ref.\cite{bars,doug}.

\vskip0.5cm

{\bf Note Added}. This paper appeared on the net simultaneously
with \cite{Bo2,Chen}, which partially overlap with it.

\acknowledgments

We would like to thank Martin Schnabl for discussions and 
stimulating suggestions.
This research was supported by the Italian MIUR 
under the program ``Teoria dei Campi, Superstringhe e Gravit\`a".

\end{document}